\documentclass[aps,prb,reprint,superscriptaddress]{revtex4-2}

\usepackage{amsmath}
\usepackage{amssymb}
\usepackage{graphicx}
\usepackage{svg}
\usepackage{placeins}
\usepackage{siunitx}
\usepackage{upgreek}
\usepackage{caption}
\usepackage[hidelinks]{hyperref}

\begin{document}

\title{High-pressure phase transitions in the quantum spin liquid candidate Na$_{2}$Co$_{2}$TeO$_{6}$ probed by Raman spectroscopy}

\author{Ihsan Ahmed Kolasseri}
\email{ihsanahmed@phys.au.dk}
\affiliation{Department of Physics and Astronomy, Aarhus University, Denmark}

\author{Maria Mei Ravnebæk}
\affiliation{Department of Chemistry, Aarhus University, Denmark}

\author{Subhadip Das}
\affiliation{Department of Physics and Astronomy, Aarhus University, Denmark}
\affiliation{Center for Quantum Devices, University of Copenhagen, Denmark}

\author{Carl Jonas Linnemann}
\affiliation{Department of Chemistry, Aarhus University, Denmark}

\author{Haidong Zhou}
\affiliation{Department of Physics and Astronomy, University of Tennessee, USA}

\author{Christian Frydendahl}
\affiliation{Department of Physics and Astronomy, Aarhus University, Denmark}

\author{Martin Bremholm}
\email{bremholm@chem.au.dk}
\affiliation{Department of Chemistry, Aarhus University, Denmark}

\author{Yong P. Chen}
\email{yongchen@phys.au.dk}
\affiliation{Department of Physics and Astronomy, Aarhus University, Denmark}
\affiliation{WPI Advanced Institute for Materials Research (AIMR) and Institute for Materials Research (IMR), Tohoku University, Japan}
\affiliation{Department of Physics and Astronomy and School of Electrical and Computer Engineering, Purdue University, USA}

\begin{abstract}
The quasi-2D magnet Na$_{2}$Co$_{2}$TeO$_{6}$ (NCTO) is a candidate for a Kitaev Quantum Spin Liquid (KQSL) state. Pressure-tuning in such materials is of interest as a potential method to tune the Kitaev exchange interactions, which are strongly dependent on bond geometry. Here we report a Raman spectroscopic study of NCTO inside a diamond anvil cell (DAC) with pressure applied up to 16.3~GPa. Based on the changes in the Raman modes, this pressure range is divided into three regions. The appearance and disappearance of several modes and changes in the polarization dependence of the representative modes, most prominently above 13.8 GPa, point to pressure-induced phase transitions in this material.
\end{abstract}

\maketitle

\section{Introduction}

The search for quantum spin liquid (QSL) states in real materials has attracted intense interest because of their potential applications in topological quantum computing and the opportunity to explore strongly entangled quantum matter. QSLs are predicted phases of matter formed by interacting quantum spins that remain disordered down to zero temperature, yet exhibit long-range quantum entanglement and fractionalized excitations \cite{feng2022detection}. An exactly solvable model of a QSL ground state for $S=1/2$ spins on a honeycomb lattice was proposed by Alexei Kitaev \cite{kitaev2006anyons}. The electron spins are coupled to the three nearest neighbors by bond-dependent Ising interactions, and the anisotropies of the interactions along the three nearest-neighbor bonds conflict with each other, giving rise to magnetic frustration \cite{takagi2019kitaev}. In real materials, strong spin-orbit coupling can create effective magnetic moments and lead to such highly frustrated anisotropic exchange interactions which can dominate over isotropic interactions and lead to a QSL state. Many magnetic insulating materials were proposed as possible KQSL candidates due to their competing spin-anisotropic exchange interactions \cite{baek2017evidence,hwan2015direct}. This led to research on 4d and 5d electron systems due to their strong spin-orbit coupling
(e.g., $\mathrm{\alpha}$-RuCl$_{3}$) \cite{banerjee2018excitations}. 

\begin{figure*}[htbp]
    \centering
    \includegraphics[width=\textwidth]{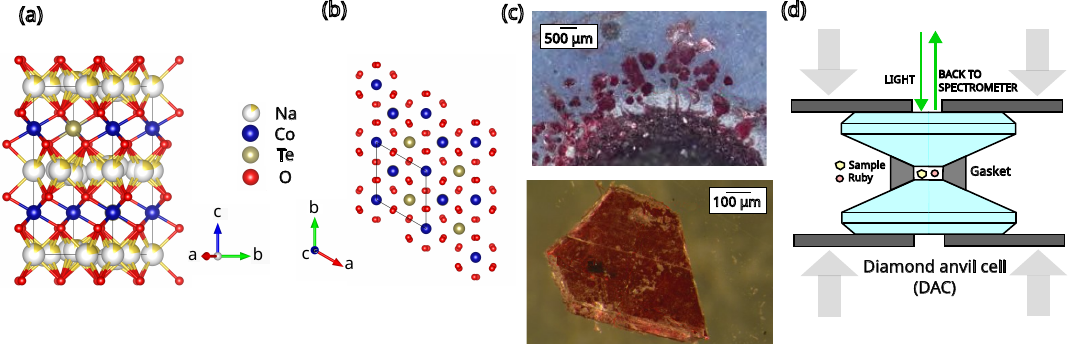}
    \caption{(a) Schematic of the crystal structure of hexagonal NCTO generated with a
    crystallographic information file from Xiao et al.\ \cite{xiao2019crystal}. The partial occupancy of the Na atoms is shown with the partial yellow color.
    (b) View of the crystal structure along the $c$ axis. Na atoms are hidden
    here to show the hexagonal arrangement of Co atoms. Each Co has 6 surrounding O atoms making a CoO$_{6}$ octahedron. In this projection the O atoms of vertically stacked octahedra overlap, so each Co appears surrounded by twelve O atoms rather than six. The black lines indicate
    the unit cell. (c) Top: `Type A' crystals from the synthesis crucible. A crystal
    with approximate size of \SI{90}{\micro\meter} was loaded into the diamond anvil cell (DAC), and was labelled as Crystal A.
    Bottom: A `Type B' crystal. The crystal size is approximately \SI{600}{\micro\meter},
    but it was cut into a piece of approximately \SI{100}{\micro\meter}, labelled as Crystal B.
    (d) Schematic diagram of a Boehler--Almax plate DAC with the sample and
    pressure sensor inside the pressure transmitting medium. The large arrows
    indicate the direction of force applied to pressurize the cell.}
    \label{fig:structure}
\end{figure*}

However, this behavior can also be seen in 3d electronic systems with weak spin-orbit
coupling \cite{kim2021spin}. The spin-orbit coupled $J_{\mathrm{eff}}=1/2$
state of Co$^{2+}$ ions with edge-sharing octahedra in materials such as NCTO facilitates orbital overlap favoring the Kitaev interaction \cite{kim2021antiferromagnetic}. Two structural phases of NCTO have been reported in the literature so far: the hexagonal phase with P6$_{3}$22 space group and the monoclinic phase with C2/m space group \cite{dufault2023introducing}. The layered
structure of hexagonal NCTO features honeycomb Co planes, shown in Fig.~\ref{fig:structure}(a) and (b). The hexagonal phase has
antiferromagnetic (AFM) ordering below 27~K with two spin reorientation
transitions. The monoclinic phase has a single AFM ordering below 9.6~K
without any spin reorientation transitions. The transition temperatures and the
magnetic ordering of these phases are very different, indicating that the spins behave differently. Inelastic neutron scattering (INS) studies show that the magnetic excitations of hexagonal NCTO can be modelled using a Kitaev-Heisenberg Hamiltonian, with calculations highlighting dominant Kitaev interactions \cite{songvilay2020kitaev,Kruger2023TripleQ,Sanders2022DominantKitaev}. Even though monoclinic NCTO shows enhanced magnetic frustration, further studies are needed to confirm Kitaev interactions in that particular phase \cite{dufault2023introducing}. Since Kitaev interactions are bond-dependent, the existence of different phases of these materials is a relevant topic to study.
Understanding and discovering different structural phases of these classes of
materials is crucial to realizing QSL states at reasonable temperatures, and to
expand the knowledge base on these materials. One way to achieve different phases is by applying pressure. For example, new phonon modes appearing at 1.1~GPa for $\mathrm{\alpha}$-RuCl$_{3}$ reflect a structural change from the monoclinic C2/m space group to the
trigonal P3$_{1}$12 space group \cite{li2019raman}. 

Here we report the first Raman study at high pressure performed on the KQSL candidate
NCTO. We applied pressures up to 16.3~GPa to investigate phase transitions in the hexagonal phase. We performed the experiment on two independently grown crystals and observed similar trends in the change of Raman modes with pressure. Polarized Raman spectroscopy at high-pressure was also performed to study the changes in these modes.

\section{Experimental details}

Two types of  single crystals (SC) with the hexagonal phase were independently synthesized
for this study. Type A crystals (Fig.~\ref{fig:structure}(c), top) were grown by a self-flux
method. For this, polycrystalline NCTO was first
prepared by a solid-state reaction. A mixture of Na$_{2}$CO$_{3}$,
Co$_{3}$O$_{4}$, and TeO$_{2}$ was prepared in a stoichiometric molar ratio
with 40\% excess Na$_{2}$CO$_{3}$. The mixture was sintered twice at
800~$^{\circ}$C for 24~hours. The resulting polycrystalline sample was then
mixed with a flux consisting of Na$_{2}$CO$_{3}$ and TeO$_{2}$ in a molar
ratio of 1:0.5:2 and heated to 900~$^{\circ}$C \cite{xiao2019crystal}. The sample was maintained at
this temperature for 30~h, after which it was cooled to 500~$^{\circ}$C at a
rate of 5~$^{\circ}$C per minute and then rapidly cooled to room temperature with the furnace turned off. The single crystal x-ray diffraction (SCXRD) of Type A crystals is shown in the Supplemental Material S1, solved in the P6$_{3}$22 space group.

The synthesis method for Type B crystals (Fig.~\ref{fig:structure}(c), bottom) follows nearly the same recipe as for the Type A crystals with small variations. For the polycrystalline sample, only 5\% excess Na$_{2}$CO$_{3}$ was used, and the mixture was sintered once at 850~$^{\circ}$C for 40~hours. Then for the self-flux method to make the single crystal, the cooling rate from 900~$^{\circ}$C to 500~$^{\circ}$C was adjusted to 3~$^{\circ}$C per hour. The structural information of Type B crystals, which are also in the hexagonal phase, can be found in \cite{lin2021field}.

High-pressure measurements were performed on both crystals, Crystal A (a SC from Type A crystals) and Crystal B (a SC from Type B crystals), using a
diamond anvil cell of the Boehler--Almax type with a diamond culet size of
\SI{400}{\micro\meter}. The schematic diagram of the DAC is shown in
Fig.~\ref{fig:structure}(d). The samples were loaded into the DAC along with a
small ruby crystal. For Crystal A, we used a 4:1 volumetric mixture of methanol
and ethanol as the pressure-transmitting medium (PTM). A stainless steel gasket with
a \SI{160}{\micro\meter} hole was used to confine the sample assembly. Ruby
fluorescence measurement was employed for pressure calibration. All Raman
measurements were carried out using a commercial Renishaw \textit{inVia} Raman
microscope. A 532~nm laser with approximately 1~mW power was used as the excitation source for Raman
measurements on the crystal and for ruby photoluminescence. A
long-working-distance 50$\times$ objective with a numerical aperture (NA) of
0.5 was employed to accommodate the DAC within the Raman experimental setup.
Polarization-dependent Raman measurements were also performed on Crystal A
at five different pressure values. A rotatable half-wave plate was inserted in
the excitation path and a linear polarizer was placed in the detection path.

For the measurements on Crystal B, an 830~nm laser with 1~mW power was used as the excitation source and a 16:3:1 volumetric mixture of methanol, ethanol, and water was used as the PTM. The rest of the parameters were the same, and polarization measurements were not performed on this crystal. 

Single crystal X-ray diffraction (SCXRD) was measured on Crystal A using a Rigaku XtaLAB Synergy-S single crystal diffractometer with Mo-K$\alpha$ radiation ($\lambda$ = 0.71073 Å). Integration of the data was performed using CrysAlisPro~\cite{cryspro} and the structure was solved using SHELXT with intrinsic phasing in Olex2 and refined using SHELXL using Least Squares minimization~\cite{cryspro,Olex2,Shelxl,Shelxt}.

\section{Results and discussion}

\begin{figure*}[htbp]
    \centering
    \includegraphics[width=\textwidth]{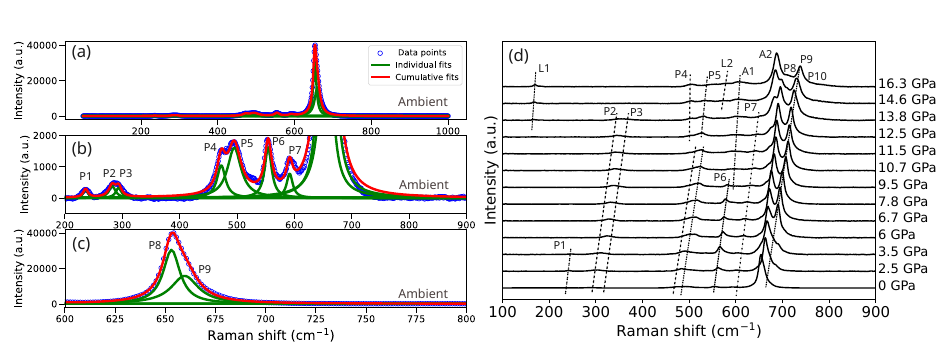}
    \caption{(a) Raman spectrum of Crystal A from 50 to 1000~cm$^{-1}$ at
    ambient pressure, fitted using Lorentzian functions. The blue points
    represent the experimental data, the red line denotes the cumulative fit,
    and the green lines indicate individual peak fits. (b) Zoomed-in view of Fig. 2(a)
    showing low-intensity Raman modes. (c) Zoomed-in view of Fig. 2(a)
    showing high-intensity Raman modes. (d) Raman spectra of Crystal A from ambient pressure (marked as 0 GPa) up to 16.3~GPa. The emergence of a new mode L1 above 13.8~GPa is clearly visible.}
    \label{fig:ambient_raman}
\end{figure*}

\begin{figure*}[htbp]
    \centering
    \includegraphics[width=\textwidth]{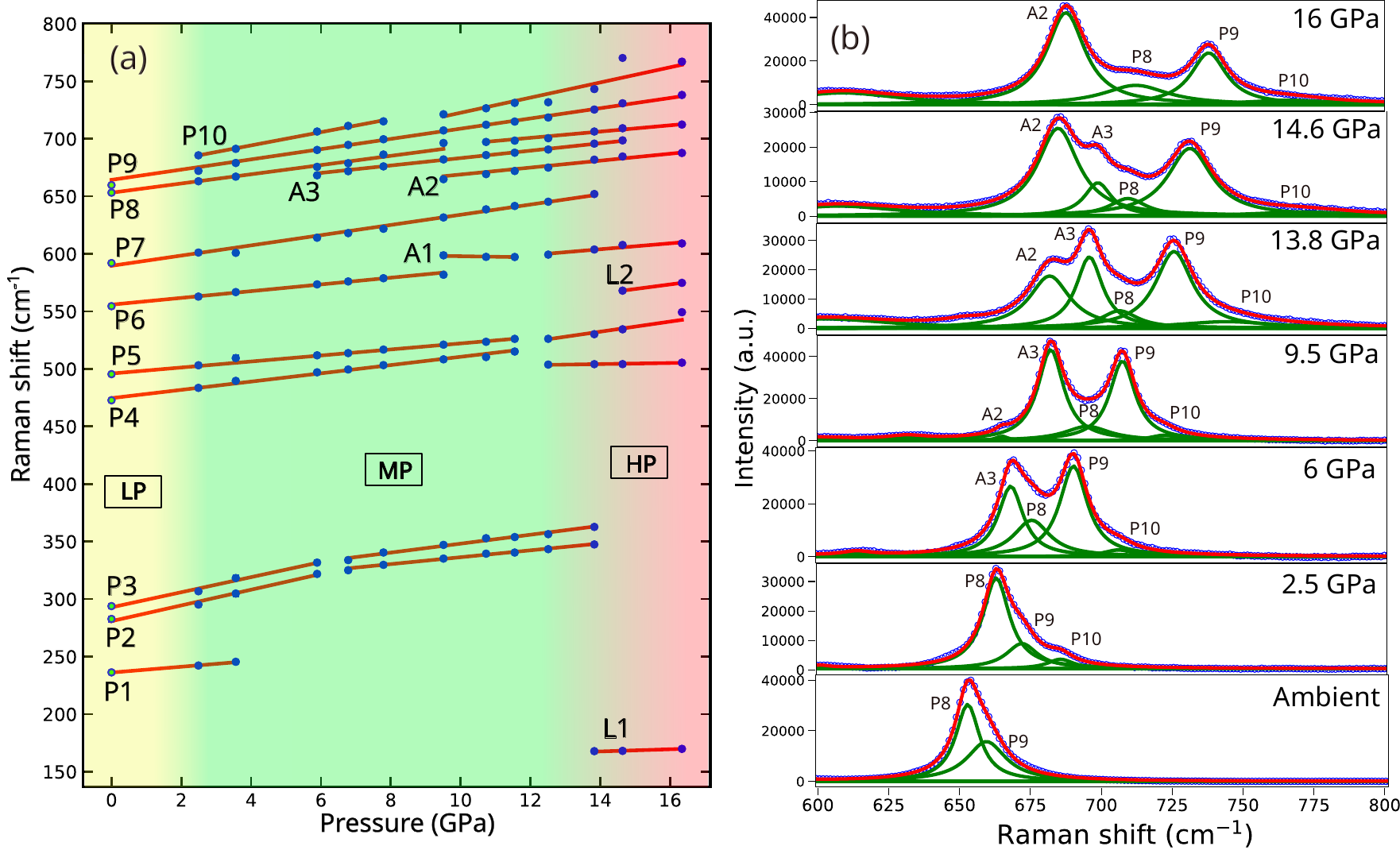}
    \caption{(a) Evolution of Raman frequencies of Crystal A with pressures up
    to 16.3~GPa. The graph is divided into three regions - LP: Low pressure regime, MP: Medium pressure regime, HP: High pressure regime. The ambient pressure points are marked in green to indicate that these data are taken from outside the DAC. (b) Evolution of modes between 600 to 800~cm$^{-1}$, up to 16.3~GPa. These high intensity peaks deconvolute into three to five peaks at various pressures.}
    \label{fig:raman_modes}
\end{figure*}

From group theory, the observable Raman-active optical branches of hexagonal NCTO with the space group P6$_{3}$22 in backscattering geometry along the c axis of the crystal are $\Gamma_{\mathrm{Raman}} = 9A_{1} +  22E_{2}$, resulting in 31 distinct frequencies \cite{chakkar2025raman}. Fig.~\ref{fig:ambient_raman}(a) shows the ambient pressure spectrum of the
hexagonal NCTO single crystal A measured outside the DAC. This is observed to be the same for
Crystal B. In the range from 50 to 1000~cm$^{-1}$ we were only able to fit nine
distinct Lorentzian peak functions. The remaining modes were not resolvable at room temperature. The obtained spectral profile matches
previously reported studies at room temperature \cite{chen2021spin,mou2024comparative,chakkar2025raman}.

Crystal A was loaded into the DAC with a starting pressure of 2.5~GPa. The Raman spectral data of Crystal A at high pressures are plotted as a waterfall
plot in Fig.~\ref{fig:ambient_raman}(d). The most interesting observations are
the changes in the high-intensity peaks between 650~cm$^{-1}$ and 780~cm$^{-1}$, and the emergence
of a new mode L1 above 13.8~GPa. Fig.~\ref{fig:raman_modes}(a) provides an
overview of the evolution of all the Raman modes with pressure for Crystal A.
The pressure dependence of each mode was fitted using a linear relation,
\begin{equation}
    \omega(P) = \omega(P_0) + C(P - P_0),
\end{equation}
where $\omega(P_0)$ is the phonon frequency at ambient pressure and $C$ is the
pressure coefficient. With increasing pressure, all Raman modes initially shift
toward higher frequencies due to the stiffening of interatomic bonds under
compression. Fig.~\ref{fig:raman_modes}(b) shows a closer look at the evolution
of the high intensity peaks. Similar analysis has been performed on Crystal B, which
can be found in Supplemental Material S3. Not all the peaks evolve in the exact same
way as in Crystal A with pressure, but both follow the same trends for the high intensity peaks and the appearance of the mode L1.

Fig.~\ref{fig:raman_modes}(a) is divided into three regions: low pressure (LP), medium pressure (MP), and high pressure (HP). The exact transition pressures between the regions differ for different NCTO crystals, but most qualitative features of the regions remain the same, hence the blurred boundaries shown. The LP region corresponds to the pressure range in which the Raman modes are similar to those at ambient pressure. For Crystal B, data were collected at more
pressure values in the LP region.  

With further increase in pressure, we enter the MP region, where more significant changes start to occur. Mode P1 is no longer distinguishable from the background signal. Within MP, there is a change in the pressure coefficients of P2 and P3. Around 9~GPa, a new peak labelled A1 emerges alongside P6, and this mode softens (negative pressure coefficient) with further increase in pressure upto 11.5 GPa. The higher intensity peaks between 650~cm$^{-1}$ and 780~cm$^{-1}$ deconvolute into three to five peaks, as clearly seen in
Fig.~\ref{fig:raman_modes}(b). At 2.5 GPa, a shoulder peak P10 emerges. At 6 GPa, A3 emerges as a high-intensity peak while P8 reduces to a shoulder peak. There is also a shoulder peak around 650~cm$^{-1}$ that becomes more prominent above background, labelled as A2, which dominates in intensity with further increase in pressure. In addition, the pressure coefficient of P8 changes.

Around 12.5~GPa, we transition to the HP region. Here, new Raman modes emerge at approximately 170~cm$^{-1}$ (L1) and 550~cm$^{-1}$ (L2). Modes P2 and P3 become indistinguishable above the relatively flat background. The pressure coefficients of P4 and P5 change as well, and mode P7
disappears. The peaks between 650~cm$^{-1}$ and 780~cm$^{-1}$ are now resolvable as only four peaks. This trend is also seen for Crystal B, where we have data at higher pressure values up to 20 GPa. 

\begin{figure}[htbp]
    \centering
    \includegraphics[width=0.7\columnwidth]{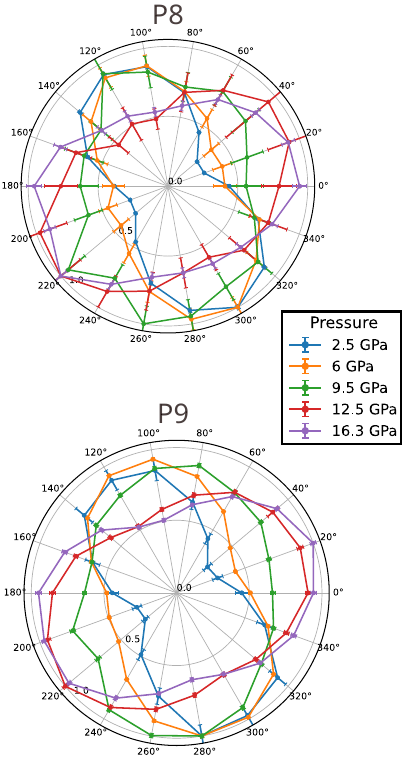}
    \caption{(a) Polar plot of normalized intensities of (a) mode P8 and (b) mode P9. The intensities of both modes are normalized by the maximum intensity of the Raman spectrum at the corresponding pressure. Zero degrees refers to the situation where the input polarization is parallel to the linear polarizer in the detection
    path.}
    \label{fig:polar}
\end{figure}

In order to further probe these regions, we performed polarized Raman spectroscopy at five different pressure values: 2.5~GPa, 6~GPa, 9.5~GPa, 12.5~GPa, and 16.3~GPa. The dataset at the lowest pressure, 2.5~GPa, is taken as the reference, and we focus on how the polarization dependence of the mode intensities changes with pressure relative to this reference. The crystal was maintained in the same orientation throughout the experiment inside the DAC (see Supplemental
Material S4 for the degree of change in sample orientation). The low-intensity
peaks in our data overlap with neighbouring peaks, which can affect the
intensities we extract from the fits. Hence in Fig.~\ref{fig:polar}, we focus on modes P8 (Fig.~\ref{fig:polar}a) and P9
(Fig.~\ref{fig:polar}b) (additional modes given in Supplemental Material S5). Both modes
show pronounced anisotropy with well-defined, two-fold patterns at 2.5 and 6~GPa (LP and MP regions), with maximum intensities near 120 and 300 degrees. The intensity minima are near 20 and 200 degrees. Above 12.5~GPa, entering the HP regime, the pattern is significantly modified where the lobes broaden, and the orientation of the intensity maxima rotates relative to the LP and MP regions. This rotation could indicate a change in the Raman tensors, and possibly in the crystal symmetry. 

To interpret the modes and their evolution with pressure, the Raman modes at ambient pressure are compared with the experimental and DFT results of Chakkar et. al. \cite{chakkar2025raman} and Mou et. al. \cite{mou2024comparative} (see Table 1 in Supplementary Material S6). Our nine resolvable modes correspond well to the reported frequencies, and mode P6 in our experiment corresponds to a phonon that has been reported to couple strongly with an underlying CoO$_{6}$ crystal field excitation. Anomalous changes in the modes P6 and A1 around 10 GPa could therefore indicate a pressure-induced change in the CoO$_{6}$ crystal field environment. To obtain information on bond specific vibrations, we compared mode assignments from Raman studies of similar honeycomb tellurates Na$_{2}$M$_{2}$TeO$_{6}$ (M=Mg, Cu, Ni) \cite{dubey2020structural, kumar2013formation} (See Figure 9 in Supplementary Material S6). The peaks between 650~cm$^{-1}$ and 780~cm$^{-1}$ correspond predominently to symmetric stretching of TeO$_{6}$  in all three compounds. In a related compound Co$_{2}$Te$_{3}$O$_{8}$, a first-order isostructural phase transition at around 14~GPa is observed, characterized by a volume collapse \cite{li2019structural}. This transition is accompanied by a relaxation of the CoO$_{6}$ octahedral distortion in
the material. This cannot be made as a one-on-one comparison, but a similar mechanism could be expected here for the material. Therefore, the changes in modes P6 and A1 and the pressure-induced switch of the polarization behaviour of the Raman modes could point to an analogous change in the CoO$_{6}$ octahedral environment. This is difficult to establish directly with the Raman data alone and could be probed by SCXRD studies in the future.

\FloatBarrier

\section{Conclusion}

This study provides evidence for pressure induced phase transitions in NCTO. It is observed that the Raman spectrum of
NCTO changes markedly at higher pressures. Above 13.8 GPa, the
emergence of the new peak L1, the disappearance of multiple modes, and the reversal of the intensity maxima of P8 and P9 from the polarization studies could point to a structural phase transition with changed symmetry. The reproducibility of these observations across two independently grown crystals
strengthens the robustness of these findings. Further structural studies, including single crystal X-ray diffraction at high pressure, can provide direct structural evidence for the transitions indicated in this work. Further theoretical studies such as density functional theories (DFT) calculations may help assign the Raman peaks to specific bond vibrations to establish more precisely the pressure evolution of specific modes and structures. Given that
NCTO hosts CoO$_{6}$ edge-sharing octahedra which are directly responsible for
mediating Kitaev interactions, this makes the high-pressure phase of NCTO a
particularly interesting target for future studies.

\begin{acknowledgments}
I.A.K., S.D. and Y.P.C. acknowledge that this work was supported by the Villum Foundation (Grant No. 25931). The work of crystal growth of Crystal A, along with M.M.R., C.J.L and M.B., was supported by The Independent Research Fund Denmark (DFF-FNU, grant no. 1026-00409B). The work at the University of Tennessee, crystal growth of Crystal B by H.Z., is supported by the National Science Foundation. C.F. was supported by the Villum Foundation (Grant No. 58634) during this work.
\end{acknowledgments}

\bibliographystyle{apsrev4-2}
\bibliography{mainclean}

\end{document}